\documentclass[12pt]{article}
\usepackage{graphicx}
\def \b{{\cal B}}
\def \bea{\begin{eqnarray}}
\def \beq{\begin{equation}}

\def \eea{\end{eqnarray}}
\def \eeq{\end{equation}}

\def \od{\overline{D}^0}

\begin{document}
\rightline{EFI 07-25}
\bigskip

\Large
\centerline{\bf Threshold effect and $\pi^\pm \psi(2S)$ peak
\footnote{To be submitted to Phys.\ Rev.\ D.}}
\normalsize
\bigskip
 
\centerline{Jonathan L. Rosner~\footnote{rosner@hep.uchicago.edu}}
\centerline{Enrico Fermi Institute and Department of Physics}
\centerline{University of Chicago, 5640 S. Ellis Avenue, Chicago, IL 60637}
 
\begin{quote}

A resonance-like structure in the $\pi^\pm \psi(2S)$ mass spectrum arising in
$B \to K \pi^\pm \psi(2S)$ has recently been reported.  It is noted that the
mass of this structure, $4433 \pm 4 \pm 1$ MeV, is not far from the
threshold for production of $D^* \overline{D}_1(2420)$.  A proposed mechanism
for production of this state is suggested, and tests are suggested.

\end{quote}

\noindent
PACS Categories: 13.25.Gv, 13.25.Hw, 14.40.Gx, 14.40.Nd

\bigskip

A wealth of charmonium states have recently been reported in $B$ meson
decays.  (For one review, see Ref.\ \cite{Eichten:2007qx}.)  Until recently,
all such states were neutral, implying the possibility of at least some
fraction of $c \bar c$ in their wave functions.  Recently, however, the Belle
Collaboration \cite{Abe:2007wg} has reported a state produced in $B \to K
\pi^\pm \psi(2S)$ in which the $\pi^\pm \psi(2S)$ system displays a
resonance-like structure with mass $M = 4433 \pm 4 \pm 1$ MeV and width
$\Gamma = 44^{+17+30}_{-13-11}$ MeV.  This would be the first observation
of a genuine tetraquark \cite{tetra} charmonium configuration.  The possibility
of easily producing such configurations in $B$ decays was noted, for example,
in Ref.\ \cite{Rosner:2003ia}.

The purpose of this Brief Report is to suggest a mechanism for production
of this state which relies upon the proximity of its mass to the
$D^*(2010) \overline{D}_1(2420)$ threshold.  S-wave thresholds appear to
be important in a wide variety of resonance-like behavior \cite{Rosner:2006vc}.
The $X(3872)$ state produced (for example) in $B \to K X$ and decaying to
$\pi^+ \pi^- J/\psi$ lies $0.6 \pm 0.6$ MeV below $D^0 \overline{D}^{*0} +$
c.c.\ threshold \cite{Cawlfield:2007dw}.  The $Y(4260)$, seen in the
radiative return reaction $e^+ e^- \to \gamma + Y(4260)$ and in a direct
$e^+ e^-$ scan, can be associated with the lowest threshold for which a
$c \bar c$ pair with $J^{PC} = 1^{--}$ can materialize into a pair of
mesons $D \overline{D}_1(2420) -$ c.c.\ in a relative S-wave
\cite{Rosner:2006vc,Close:2004ip}.

\begin{figure}[h]
\begin{center}
\includegraphics[height=2.7in]{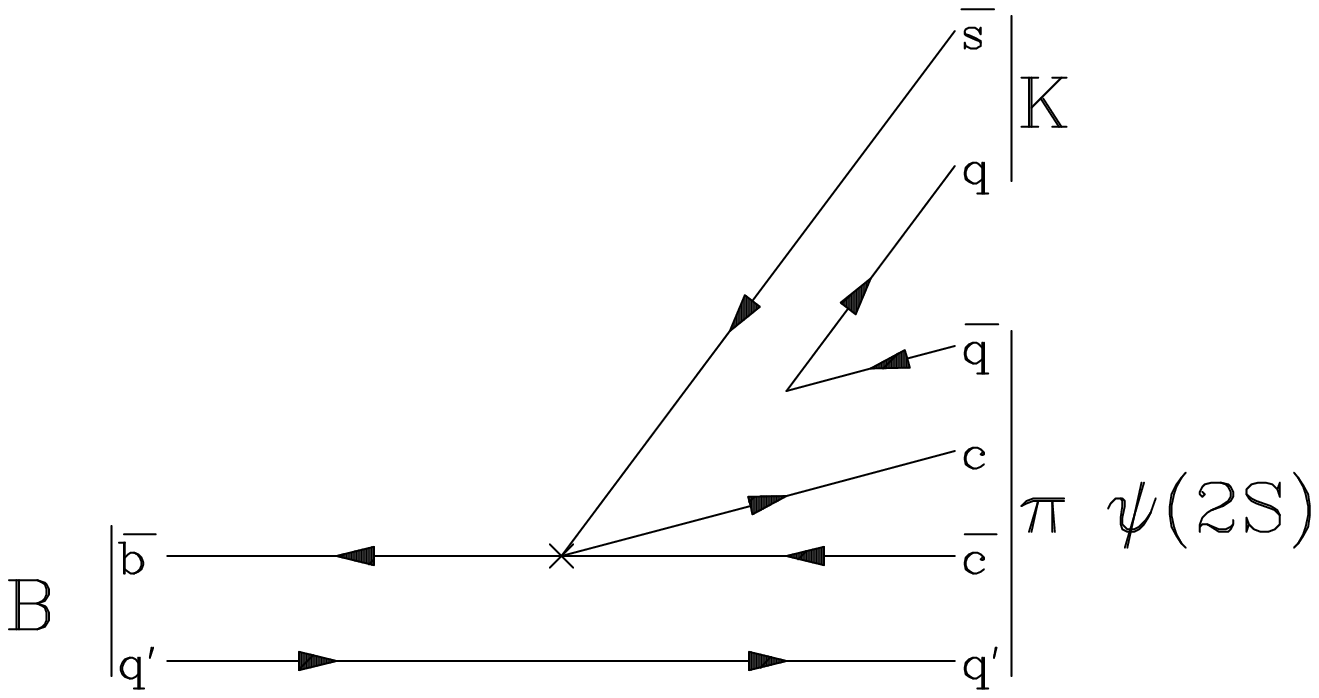}
\end{center}
\caption{Diagram illustrating the production of a $\pi \psi(2S)$ state
in $B$ decays.  The weak subprocess $\bar b \to \bar c c \bar s$ is labeled
by $\times$.
\label{fig:p2s}}
\end{figure}

The production mechanism we suggest for the $\pi^\pm \psi(2S)$ resonance-like
state is based on the diagram of Fig.\ \ref{fig:p2s}.  The different charge
states that can be involved in this process are summarized in Table
\ref{tab:p2s}.

% This is Figure 1

% This is Table I
\begin{table}
\caption{Possible charge states for production of a $\pi \psi(2S)$ state
in $B$ decays.
\label{tab:p2s}}
\begin{center}
\begin{tabular}{c c c c c} \hline \hline
$q$ & $q'$ &  $B$  &  $K$  & $Z(4430) \to$ \\ \hline
$u$ & $d$  & $B^0$ & $K^+$ & $\pi^- \psi(2S)$ \\
$d$ & $u$  & $B^+$ & $K^0$ & $\pi^+ \psi(2S)$ \\
$u$ & $u$  & $B^+$ & $K^+$ & $\pi^0 \psi(2S)$ \\
$d$ & $d$  & $B^0$ & $K^0$ & $\pi^0 \psi(2S)$ \\ \hline \hline
\end{tabular}
\end{center}
\end{table}

The quarks $q$ and $q'$ are independent.  Isospin invariance implies
$\b[B^0 \to K^+ \pi^- \psi(2S)] = 2 \b[B^0 \to K^0 \pi^0 \psi(2S)]$ and
$\b[B^+ \to K^0 \pi^+ \psi(2S)] = 2 \b[B^+ \to K^+ \pi^0 \psi(2S)]$.

The proposed mechanism operates by the production of an anti-charmed
meson $\bar c q'$ and a charmed meson $c \bar q$ which then rescatter into
$c \bar c = \psi(2S)$ and $ q' \bar q = \pi$.  A key feature of the data
not answered by the present mechanism is why rescattering into $J/\psi
\pi$ is not observed.  Perhaps the rescattering process is enhanced when the
Q-values of the two sides are more nearly equal.  The additional Q-value
available in rescattering into states containing $J/\psi$ may favor higher
pion multiplicities, e.g., $3 \pi J/\psi$ or even $5 \pi J/\psi$, over
$\pi J/\psi$ \cite{Snyder}.  [Here we have assumed a definite G-parity
$G(Z) = +$.]

The $\bar c q'$ meson can be either $\overline{D}_1(2420)$ (the narrow
P-wave charmed meson decaying to $\overline{D}^* \pi$) or $\overline{D}^*
(2010)$ (the vector meson state decaying to $\overline{D}\pi$).  The $c \bar q$
meson would then correspondingly be $D^*(2010)$ or $D_1(2420)$.  In
either case, the final state $D^* \overline{D}^* \pi$ should be visible,
with a Dalitz plot showing a strong $\overline{D}_1(2420)$ and/or $D(2420)$
band.  Which band is populated can shed light on details of the decay
mechanism, such as whether relative orbital angular momentum of zero or one
is favored between the $\bar c$ and the $q'$ in Fig.\ \ref{fig:p2s}.

The S-wave states of $D^*(2010) + \overline{D}_1(2420)$ can have spin-parity
$J^P = 0^-,1^-,2^-$.  A $0^-$ or $1^-$ state would decay to $\pi \psi(2S)$ via
a P-wave, while either P-wave or F-wave decay would be allowed for $2^-$.
The calculation of acceptance in Ref.\ \cite{Abe:2007wg} assumed a relative
S-wave between $\pi^\pm$ and $\psi(2S)$.  The rather low Q-value for
the decay $B \to K Z(4430)$ likely favors a low angular momentum $\ell$
between $K$ and $Z$.  A low spin $J(Z)$ is then favored since one must have
$J(Z)=\ell$ in this decay.  For $J^P(Z) = 0^-$, the polarization vector of the
$\psi(2S)$ in $Z \to \pi \psi(2S)$ must be parallel to the direction of
the recoil $\pi$ in the rest frame of the $\psi(2S)$.  If the polarization
of the $J/\psi$ follows that of the $\psi(2S)$ (a good approximation), the
leptons in $J/\psi \to \ell^+ \ell^-$ will have a $\sin^2 \theta$ distribution
with respect to the recoil $\pi$ momentum.

If the $q \bar q$ pair in Fig.\ \ref{fig:p2s} is $s \bar s$ rather than $u \bar
u$ or $d \bar d$, one will have final states such as $\phi D_s^{(*)} D^{(*)}$
or even (barely) $\phi D_s(2317) D$ \cite{Snyder}.  The charm-anticharm pair
could then rescatter into $K J/\psi$ or (for $D_s D$) $K \psi(2S)$.  The
decay $B^+ \to K^+ \phi J/\psi$ has been observed with a branching
ratio of $(5.2 \pm 1.7) \times 10^{-5}$ (average of Ref.\ \cite{PDG07},
based on Refs.\ \cite{Jessop:1999cr} and \cite{Aubert:2003ii}), and should be
examined for bumps in the $K^+ J/\psi$ spectrum.

An anaglogue in charm decays, in which one would search for a $\phi \pi^-$
resonance, would be the Cabibbo-suppressed decay $D^0 \to K^+ K^- \pi^+
\pi^-$ \cite{Snyder}.  If the mechanism of Fig.\ \ref{fig:p2s} is responsible
for a resonance through rescattering from a $K^{(*)} \bar K^{(*)}$ state,
$D^0$ decays will yield a $\phi \pi^-$ resonance while $\od$ decays will
yield a $\phi \pi^+$ resonance.

An alternative mechanism for production of a $c \bar c \pi$ state, distinct
from that shown in Fig.\ \ref{fig:p2s}, would involve a $\bar b \to \bar s$
penguin transition, leading to a similar diagram but with the $c \bar c$
pair produced from the vacuum rather than at the weak vertex.  The presence
of a signal in $\pi \psi(2S)$ and its absence in $\pi J/\psi$ would be even
more puzzling in this picture.  Moreover, the large product branching ratio
\cite{Abe:2007wg},
\beq
\b[B \to K Z(4430)] \times \b[Z(4430) \to \pi^+ \psi(2S)] = (4.1 \pm 1.0 \pm
1.3) \times 10^{-5}~,
\eeq
is larger than most $\bar b \to \bar s$ penguin-dominated processes {\it
without} charmed pair production, so this alternative mechanism is highly
unlikely to account for the observed signal.  A similar statement applies
to the case of the weak subprocess $\bar b \to \bar u u \bar s$ accompanied
by charmed pair production from the vacuum, as this subprocess is even
weaker than the $\bar b \to \bar s$ penguin process.
\bigskip

[Note added:  subsequently to this work, a proposal appeared \cite{Maiani:2007}
that the $Z(4430)$, whose neutral member has charge conjugation eigenvalue
$C=-$, is a tetraquark state representing a radial excitation of
an as-yet-unseen $C=-$ state not far in mass from the $X(3872)$.  (The
$X(3872)$ is identified as having $C=+1$ through its decay to $\gamma J/\psi$
\cite{Aubert:2006,Abe:2005}.)  Even more recently, a proposal similar to ours
\cite{Meng:2007} accounts for the apparent enhancement of the ratio
$\Gamma[Z(4430) \to \pi \psi(2S)]/\Gamma[Z(4430 \to \pi J/\psi]$ via a
rescattering model based on charm exchange, and concludes that $J^P[Z(4430)] =
1^-$ is favored.]
\bigskip

I thank Vera Luth, Luciano Maiani, and Art Snyder for discussions.   Part of
this work was performed at the Aspen Center for Physics.  This work was
supported in part by the United States Department of Energy through Grant No.\
DE FG02 90ER40560.

% Journal and other miscellaneous abbreviations for references
% Phys. Rev. D format
\def \ajp#1#2#3{Am.\ J. Phys.\ {\bf#1}, #2 (#3)}
\def \apny#1#2#3{Ann.\ Phys.\ (N.Y.) {\bf#1}, #2 (#3)}
\def \app#1#2#3{Acta Phys.\ Polonica {\bf#1}, #2 (#3)}
\def \arnps#1#2#3{Ann.\ Rev.\ Nucl.\ Part.\ Sci.\ {\bf#1}, #2 (#3)}
\def \art{and references therein}
\def \cmts#1#2#3{Comments on Nucl.\ Part.\ Phys.\ {\bf#1}, #2 (#3)}
\def \cn{Collaboration}
\def \cp89{{\it CP Violation,} edited by C. Jarlskog (World Scientific,
Singapore, 1989)}
\def \efi{Enrico Fermi Institute Report No.\ }
\def \epjc#1#2#3{Eur.\ Phys.\ J. C {\bf#1}, #2 (#3)}
\def \f79{{\it Proceedings of the 1979 International Symposium on Lepton and
Photon Interactions at High Energies,} Fermilab, August 23-29, 1979, ed. by
T. B. W. Kirk and H. D. I. Abarbanel (Fermi National Accelerator Laboratory,
Batavia, IL, 1979}
\def \hb87{{\it Proceeding of the 1987 International Symposium on Lepton and
Photon Interactions at High Energies,} Hamburg, 1987, ed. by W. Bartel
and R. R\"uckl (Nucl.\ Phys.\ B, Proc.\ Suppl., vol.\ 3) (North-Holland,
Amsterdam, 1988)}
\def \ib{{\it ibid.}~}
\def \ibj#1#2#3{~{\bf#1}, #2 (#3)}
\def \ichep72{{\it Proceedings of the XVI International Conference on High
Energy Physics}, Chicago and Batavia, Illinois, Sept. 6 -- 13, 1972,
edited by J. D. Jackson, A. Roberts, and R. Donaldson (Fermilab, Batavia,
IL, 1972)}
\def \ijmpa#1#2#3{Int.\ J.\ Mod.\ Phys.\ A {\bf#1}, #2 (#3)}
\def \ite{{\it et al.}}
\def \jhep#1#2#3{JHEP {\bf#1}, #2 (#3)}
\def \jpb#1#2#3{J.\ Phys.\ B {\bf#1}, #2 (#3)}
\def \lg{{\it Proceedings of the XIXth International Symposium on
Lepton and Photon Interactions,} Stanford, California, August 9--14 1999,
edited by J. Jaros and M. Peskin (World Scientific, Singapore, 2000)}
\def \lkl87{{\it Selected Topics in Electroweak Interactions} (Proceedings of
the Second Lake Louise Institute on New Frontiers in Particle Physics, 15 --
21 February, 1987), edited by J. M. Cameron \ite~(World Scientific, Singapore,
1987)}
\def \kdvs#1#2#3{{Kong.\ Danske Vid.\ Selsk., Matt-fys.\ Medd.} {\bf #1},
No.\ #2 (#3)}
\def \ky85{{\it Proceedings of the International Symposium on Lepton and
Photon Interactions at High Energy,} Kyoto, Aug.~19-24, 1985, edited by M.
Konuma and K. Takahashi (Kyoto Univ., Kyoto, 1985)}
\def \mpla#1#2#3{Mod.\ Phys.\ Lett.\ A {\bf#1}, #2 (#3)}
\def \nat#1#2#3{Nature {\bf#1}, #2 (#3)}
\def \nc#1#2#3{Nuovo Cim.\ {\bf#1}, #2 (#3)}
\def \nima#1#2#3{Nucl.\ Instr.\ Meth. A {\bf#1}, #2 (#3)}
\def \np#1#2#3{Nucl.\ Phys.\ {\bf#1}, #2 (#3)}
\def \npbps#1#2#3{Nucl.\ Phys.\ B Proc.\ Suppl.\ {\bf#1}, #2 (#3)}
\def \os{XXX International Conference on High Energy Physics, Osaka, Japan,
July 27 -- August 2, 2000}
\def \PDG{Particle Data Group, K. Hagiwara \ite, \prd{66}{010001}{2002}}
\def \pisma#1#2#3#4{Pis'ma Zh.\ Eksp.\ Teor.\ Fiz.\ {\bf#1}, #2 (#3) [JETP
Lett.\ {\bf#1}, #4 (#3)]}
\def \pl#1#2#3{Phys.\ Lett.\ {\bf#1}, #2 (#3)}
\def \pla#1#2#3{Phys.\ Lett.\ A {\bf#1}, #2 (#3)}
\def \plb#1#2#3{Phys.\ Lett.\ B {\bf#1}, #2 (#3)}
\def \pr#1#2#3{Phys.\ Rev.\ {\bf#1}, #2 (#3)}
\def \prc#1#2#3{Phys.\ Rev.\ C {\bf#1}, #2 (#3)}
\def \prd#1#2#3{Phys.\ Rev.\ D {\bf#1}, #2 (#3)}
\def \prl#1#2#3{Phys.\ Rev.\ Lett.\ {\bf#1}, #2 (#3)}
\def \prp#1#2#3{Phys.\ Rep.\ {\bf#1}, #2 (#3)}
\def \ptp#1#2#3{Prog.\ Theor.\ Phys.\ {\bf#1}, #2 (#3)}
\def \rmp#1#2#3{Rev.\ Mod.\ Phys.\ {\bf#1}, #2 (#3)}
\def \rp#1{~~~~~\ldots\ldots{\rm rp~}{#1}~~~~~}
\def \rpp#1#2#3{Rep.\ Prog.\ Phys.\ {\bf#1}, #2 (#3)}
\def \sing{{\it Proceedings of the 25th International Conference on High Energy
Physics, Singapore, Aug. 2--8, 1990}, edited by. K. K. Phua and Y. Yamaguchi
(Southeast Asia Physics Association, 1991)}
\def \slc87{{\it Proceedings of the Salt Lake City Meeting} (Division of
Particles and Fields, American Physical Society, Salt Lake City, Utah, 1987),
ed. by C. DeTar and J. S. Ball (World Scientific, Singapore, 1987)}
\def \slac89{{\it Proceedings of the XIVth International Symposium on
Lepton and Photon Interactions,} Stanford, California, 1989, edited by M.
Riordan (World Scientific, Singapore, 1990)}
\def \smass82{{\it Proceedings of the 1982 DPF Summer Study on Elementary
Particle Physics and Future Facilities}, Snowmass, Colorado, edited by R.
Donaldson, R. Gustafson, and F. Paige (World Scientific, Singapore, 1982)}
\def \smass90{{\it Research Directions for the Decade} (Proceedings of the
1990 Summer Study on High Energy Physics, June 25--July 13, Snowmass, Colorado),
edited by E. L. Berger (World Scientific, Singapore, 1992)}
\def \tasi{{\it Testing the Standard Model} (Proceedings of the 1990
Theoretical Advanced Study Institute in Elementary Particle Physics, Boulder,
Colorado, 3--27 June, 1990), edited by M. Cveti\v{c} and P. Langacker
(World Scientific, Singapore, 1991)}
\def \yaf#1#2#3#4{Yad.\ Fiz.\ {\bf#1}, #2 (#3) [Sov.\ J.\ Nucl.\ Phys.\
{\bf #1}, #4 (#3)]}
\def \zhetf#1#2#3#4#5#6{Zh.\ Eksp.\ Teor.\ Fiz.\ {\bf #1}, #2 (#3) [Sov.\
Phys.\ - JETP {\bf #4}, #5 (#6)]}
\def \zpc#1#2#3{Zeit.\ Phys.\ C {\bf#1}, #2 (#3)}
\def \zpd#1#2#3{Zeit.\ Phys.\ D {\bf#1}, #2 (#3)}

\end{document}